\documentclass[epj]{svjour}
\usepackage{latexsym}
\usepackage{graphicx}
\usepackage{amsmath}
\usepackage{amssymb}
\usepackage{xcolor}
\usepackage{ulem}

\begin{document}

\title{ Coherent control of single photons in the cross resonator arrays via
the dark state mechanism}
\author{ T. Tian\inst{1} \and D. Z. Xu\inst{1} \and T. Y. Zheng\inst{2} \and %
C. P. Sun\inst{3}\thanks{suncp@itp.ac.cn}}
\institute{Key laboratory of Frontiers in theoretical Physics, Institute of Theoretical
Physics, Chinese Academy of Sciences, Beijing, 100080, China \and School of
Physics, Northeast Normal University, Changchun, 130024, China \and Beijing
Computational Science Research Center, Beijing 100084, China}
\date{ Received: date / Revised version: date}

\abstract{ We study the single photon transfer in a hybrid system where the normal
modes of two coupled resonator arrays interact with two transition arms of a
$\Lambda$-type atom localized in the intersectional resonator. It is found
that, due to the Fano-Feshbach effect based on the dark state of the $\Lambda $-type atom, the photon transfer in one array can be well controlled
by the bound state of the photon in the other array. This conceptual setup could
be implemented in some practical cavity QED system to realize a quantum
switch for single photon.
}
\maketitle

\section{ Introduction}

\label{introduction}

In quantum information physics and technology, photon plays an important
role since it can robustly transfer information over long distance as a
flying qubit in free space. High-fidelity transfer of an independently
prepared quantum state from photons onto atomic ensemble has been
experimentally feasible\cite{bju}. Most recently, the study of confined
photon in low dimension structure, such as the coupled resonator array
(CRA), is attracting more and more attentions\cite{mjh1,adg,mjh2,mjh3,lzh,lzh2}. The nonlinear dispersion relation of CRA
system can result in single photon quasi-bound states \cite{lzh,zrg,hdo},
which can be applied to realize information storage and coherent control of
single photon transmission in a hybrid system. Some special atomic mediums
enhance the nonlinearity of the resonator hence are capable for
demonstrating the photon blockade phenomenon\cite{aim,kmg,kmb}.

Moreover, one-dimension wave guide constructed by CRA with atom embedded in
can realize well controllable photon transport. The two-level atom in the
wave guide acts as a perfect mirror for the light field at resonance\cite%
{hdo}. To realize a better tunable mirror, people use the three-level atom
instead of the two-level one, thus the electromagnetically induced
transparency(EIT) mechanism can be utilized to control the behavior of the
probe photon by a classical control light beam\cite{zrg,mjw}. However, we
prefer a full quantum network without introducing any classical element.
This consideration motivates us to discuss the Fano-Feshbach resonate in
CRA system, which is analog of the Fano resonance in ionize process of atom
system\cite{uf} or the Feshbach resonate used in the cold atom system for
controlling the interaction strength\cite{hfe,eti}. In our consideration, if
a photon bound state is formed in one transfer channel, the transmission
feature of the photon in another channel is greatly influenced\cite{dzx}.

In this paper, we study the coherent transport of a single photon in two
crossed CRAs with a $\Lambda$-type atom embedded in the intersectional
resonator. We use the discrete coordinates method\cite{lzh} to calculate the
transmission and reflection coefficients of the incident photon. We find
that the photon incidenting in one array is perfectly reflected when it
resonates with single photon bound states in the other array, namely, we use
the Feshbach resonance mechanism to control the transmission of single
photon in this two-channel CRA system. Under two-photon resonate condition,
we find that when the incident photon is perfectly transmitted or reflected,
its wave function has the maximum overlap with the dark or bright states in
the intersectional resonator. This implies the EIT mechanism\cite{seh1,seh2}
intrinsiclly exists in our system.

This paper is organized as follows: In Sec.~\ref{sec2}, we present the model
Hamiltonian for single photon scattered by a $\Lambda$-type atom in two
crossed CRAs system. In Sec.~\ref{sec3}, we study single photon scattering
process, and the transmission coefficient is obtained by discrete
coordinates scattering equations. In Sec.~\ref{sec4}, three nontrivial cases
that the photon is totally reflected or transmitted are well studied. In
Sec.~\ref{sec5}, we show the role of the dark state mechanism in the
controlling of photon transmission. The conclusion and physical realization is given in Sec.~\ref%
{sec6}.

\section{ Model setup}

\label{sec2} 
\begin{figure}[tbp]
\includegraphics[bb=34 225 438 638, width=8cm]{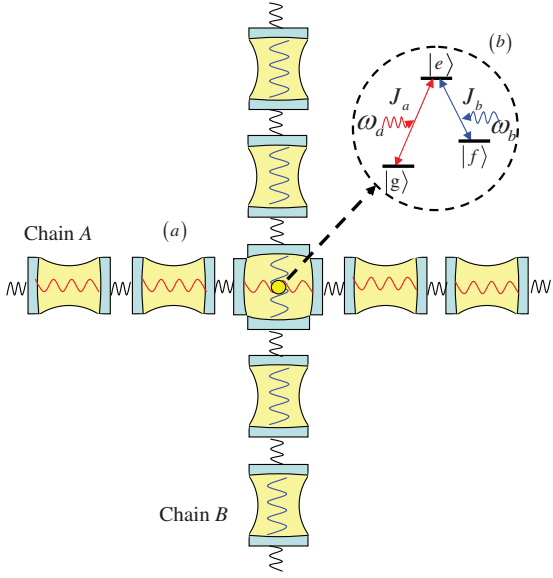} \vspace{0.5cm}
\caption{(Color online)Schematic configuration for single photons transition
in cross CRAs. (a)Two cross CRAs in which a $\Lambda$-type atom is located
inside the intersectional resonator. (b)In the intersectional resonator
there are two orthogonal resonator modes with different frequencies which
interact with two transitions of the atom respectively. We define the
horizontal arrays as chain \textit{A}, the vertical arrays as chain
\textit{B}.}
\label{fig1}
\end{figure}

We consider two crossed CRAs with a $\Lambda$-type atom system,
the horizontal(vertical) CRA is named as the chain \textit{A}   (\textit{B}) with
resonators' eigen modes $\omega_{a}$($\omega_{b}$). The system is
schematically illustrated in Fig.~\ref{fig1}.

\vspace{0cm}

The tight-binding Hamiltonian of the crossed CRAs reads

\begin{eqnarray}
H_{C} & = &
\omega_{a}\sum_{m}a_{m}^{\dagger}a_{m}+\omega_{b}\sum_{n}b_{n}^{\dagger}b_{n}
\notag \\
& &
-\xi_{a}\sum_{m}\left(a_{m}^{\dagger}a_{m+1}+a_{m+1}^{\dagger}a_{m}\right)
\notag \\
& &
-\xi_{b}\sum_{n}\left(b_{n}^{\dagger}b_{n+1}+b_{n+1}^{\dagger}b_{n}\right),
\end{eqnarray}
where $a_{m}$ ($b_{n}$) is the photon annihilation operator of chain
\textit{A} (\textit{B}), $m $ and $n$ are all integers indicating the positions
of the resonators corresponding to the chain \textit{A} and the chain \textit{B}%
. $\xi_{a}$ and $\xi_{b}$ are the hopping energies between two
nearest-neighbor resonators of the two chains, respectively.

The intersectional resonator which is labeled by $0$ is supposed to be able
to support two CRA modes. The $\Lambda$-type atom with a ground state $%
|g\rangle$, a metastable state $|f\rangle$ and an excited state $|e\rangle$
is placed in the intersectional resonator. We choose the resonators' mode
such that $\omega_{a}$ only couples to the transition between $|g\rangle$
and $|e\rangle$ while the mode $\omega_{b} $ only couples to the transition
between $|f\rangle$ and $|e\rangle$, $J_{a}$ and $J_{b}$ are the coupling
strengths respectively. Under the rotating wave approximation, the
atom-photon interaction is modeled as a Jaynes-Cummings Hamiltonian,

\begin{eqnarray}
H_{I} &=&\varepsilon _{e}\left\vert e\right\rangle \left\langle e\right\vert
+\varepsilon _{f}\left\vert f\right\rangle \left\langle f\right\vert
+J_{a}\left( a_{0}\left\vert e\right\rangle \left\langle g\right\vert
+H.c\right)  \notag \\
&&+J_{b}\left( b_{0}\left\vert e\right\rangle \left\langle f\right\vert
+H.c\right) .
\end{eqnarray}

We choose the energy of the ground state as zero, $\varepsilon _{e}$ and $%
\varepsilon _{f}$ are the energies of the excited state and the metastable
state respectively. Thereafter, we set $\hbar =1$.

\section{Single photons scattering}

\label{sec3} To explore the single photon scattering behavior in this model,
we consider a single photon incidents into the vertical resonator array
chain \textit{B} from the resonator at $-\infty$. Because the total excitation
number $N=|e \rangle\langle
e|+\sum_{m}a_{m}^{\dagger}a_{m}+\sum_{n}b_{n}^{\dagger}b_{n}$ is conserved. The eigen state of the total
Hamiltonian $H=H_{C}+H_{I}$ can be expressed as

\begin{equation}
\left\vert E\right\rangle =\sum_{m}u_{g}\left( m\right) \left\vert
m,g\right\rangle +\sum_{n}u_{f}\left( n\right) \left\vert n,f\right\rangle
+u_{e}\left\vert \phi,e\right\rangle ,  \label{state}
\end{equation}%
within the single excitation subspace. Where $\left\vert m,g\right\rangle
\equiv \left\vert m\right\rangle \otimes \left\vert g\right\rangle $ ($%
\left\vert n,f\right\rangle \equiv \left\vert n\right\rangle \otimes
\left\vert f\right\rangle $) is the state with one photon in the $m$th ($n$%
th) resonator of the chain \textit{A} (\textit{B}) while the atom in the ground
(metastable) state, $\left\vert \phi,e\right\rangle$ is the state with no
photons in the CRAs and the atom in the excited state. $u_{g}\left( m\right)
$, $u_{f}\left( n\right) $, and $u_{e}$ are the probability amplitudes for
the corresponding states. According to the stationary Schr\"{o}dinger
equation $H\left\vert E\right\rangle =E\left\vert E\right\rangle $, we
eliminate $u_{e}$ and obtain the equations for $u_{g}\left( m\right) $ and $%
u_{f}\left( n\right) $ as:

\begin{eqnarray}
\left[E-\omega_{a}+V_{a}\left(m\right)\right]u_{g}\left(m\right)
+V\left(m\right)u_{f}\left(0\right)  \notag \\
= -\xi_{a}\left[u_{g}\left(m+1\right)+u_{g}\left(m-1\right)\right],
\label{fc5} \\
\left[E-\omega_{b}-\varepsilon _{f}+V_{b}\left(n\right)\right]%
u_{f}\left(n\right) +V\left(n\right)u_{g}\left(0\right)  \notag \\
= -\xi_{b}\left[u_{f}\left(n+1\right)+u_{f}\left(n-1\right)\right].
\label{fc6}
\end{eqnarray}
Here, we define the effective potentials as

\begin{eqnarray}
V_{a\left( b\right) }\left( i\right) &\equiv&\frac{J_{a\left( b\right) }^{2}}{%
E-\varepsilon _{e}}\delta _{i,0},  \label{1} \\
V\left( i\right) &\equiv&\frac{J_{a}J_{b}}{E-\varepsilon _{e}}\delta _{i,0}.
\label{3}
\end{eqnarray}

The strengths of the $\delta $-type potentials $V_{a\left( b\right) }$ and $%
V $ are related to the hopping strength $J_{a\left( b\right) }$, the
excitation energy $\varepsilon _{e}$ of the atom, and especially the energy $%
E$ of the incident photon itself. We assume and the probability amplitudes $u_{g}\left( m\right)$ in chain \textit{A} and $u_{f}\left(
n\right)$ in chain \textit{B} have the plane-wave solutions,

\begin{equation}
u_{g}\left( m\right) =\left\{
\begin{array}{c}
Ae^{ikm}\text{ \ \ \ }m<0, \\
Ae^{-ikm}\text{ \ \ }m>0,%
\end{array}%
\right.  \label{jie2}
\end{equation}%
and
\begin{equation}
u_{f}\left( n\right) =\left\{
\begin{array}{c}
e^{ik^{\prime }n}+re^{-ik^{\prime }n}\text{ \ \ }n<0, \\
se^{ik^{\prime }n}\text{ \ \ \ \ \ \ \ \ \ \ \ \ \ }n>0,%
\end{array}%
\right.  \label{jie1}
\end{equation}%
where $A$ is the normalization constant, $r$ and $s$ are reflection and
transmission coefficients of the photon in chain \textit{B}. $k$ and $%
k^{\prime }$ are the wave vectors of the two chains.

Substitute Eq.$\left(\ref{jie2}\right)$ and Eq.$\left(\ref{jie1}\right)$ into
the scattering equations Eqs.$\left(\ref{fc5}\right)$ and $\left(\ref{fc6}%
\right)$, when $m,n\neq0$, we obtain the dispersion relations for two chains
as:

\begin{align}
E& =\omega _{a}-2\xi _{a}\cos k,  \label{scatter2} \\
E& =\omega _{b}+\varepsilon _{f}-2\xi _{b}\cos k^{\prime }.  \label{scatter1}
\end{align}

In the intersectional resonator, the continuous condition $%
u_{f}\left(0^{+}\right)=u_{f}\left(0^{-}\right)$ leads to

\begin{equation}
1+r=s.  \label{lianxu}
\end{equation}

We solve the scattering equations Eq.$\left( \ref{fc5}\right) $ and $\left( %
\ref{fc6}\right) $ for the intersectional resonator with the help of Eq.~(%
\ref{scatter2}-\ref{lianxu}) and obtain the transmission amplitude $s$ as

\begin{equation}
s=\frac{i\kappa \left( E\right) }{i\kappa \left( E\right) +\frac{%
J_{a}^{2}J_{b}^{2}}{J_{a}^{2}\left( E-\varepsilon _{e}\right) +\left(
E-\varepsilon _{e}\right) ^{2}\zeta \left( E\right) }-\frac{J_{b}^{2}}{%
\left( E-\varepsilon _{e}\right) }}  \label{s}
\end{equation}%
with $\kappa \left( E\right) $ is defined as
\begin{equation}
\kappa \left( E\right) \equiv \sqrt{4\xi _{b}^{2}-\left( E-\omega
_{b}-\varepsilon _{f}\right) ^{2}}
\end{equation}%
and
\begin{equation}
\zeta \left( E\right) =\left\{
\begin{array}{c}
\sqrt{f\left( E\right) },\text{ \ \ \ \ }E\in \left[ \omega
_{b}+\varepsilon _{f}-2\xi _{b},\omega _{a}-2\xi _{a}\right]  \\
i\sqrt{f\left( E\right) },E\in \left[ \omega _{a}-2\xi _{a},\omega
_{a}+2\xi _{a}\right]  \\
-\sqrt{-f\left( E\right) }.E\in \left[ \omega _{a}+2\xi _{a},\omega
_{b}+\varepsilon _{f}+2\xi _{b}\right]
\end{array}%
\right.
\end{equation}%
where we introduce the notation $f\left( E\right) =4\xi _{a}^{2}-\left(
E-\omega _{a}\right) ^{2}$.

\section{Fano-Feshbach resonate effect}

\label{sec4}

In the above section we have obtained the single photons transmission
amplitude in chain \textit{B}. From Eq.~(\ref{scatter2}) and Eq.~(\ref%
{scatter1}), the energy spectrums for single photons in chain \textit{A} and
chain \textit{B} have band structures. Due to the interaction with the $\Lambda
$-type atom, there are also two isolated bound states levels outside the
band. In chain \textit{A}, if the wave vector $k$ has a negative imaginary
part the photon wave function will decay with the distance from the
intersection resonator. We call this state a single photon bound state. The
complex wave vector $k $ with negative imaginary part corresponds to the
photon bound states. Single photon bound state in one-dimension CRA is
discussed in Appendix.

\begin{figure}[tbp]
\resizebox{0.45\textwidth}{!}{
\includegraphics{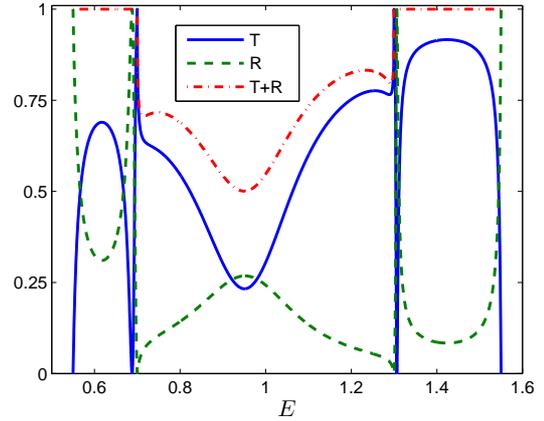}
}
\caption{(Color online) The photon transmission rate $T=\left\vert
s\right\vert ^{2}$(blue and solid line), reflection rate $R=\left\vert
r\right\vert ^{2}$(green and dashed line) and $T+R$ (red and dotted dashed
line) against the incident energy $E$. Parameters of the system are set as
follows: $\protect\xi _{a}=0.15$,$\protect\xi _{b}=0.25$,$J_{a}=0.15$,$%
J_{b}=0.2$,$\protect\omega _{b}=0.9$,$\protect\varepsilon _{e}=0.95$,$%
\protect\varepsilon _{f}=0.15$. All parameters are in units of $\protect%
\omega _{a}$.}
\label{fig2}
\end{figure}

\begin{figure*}[ht]
\begin{center}
\includegraphics{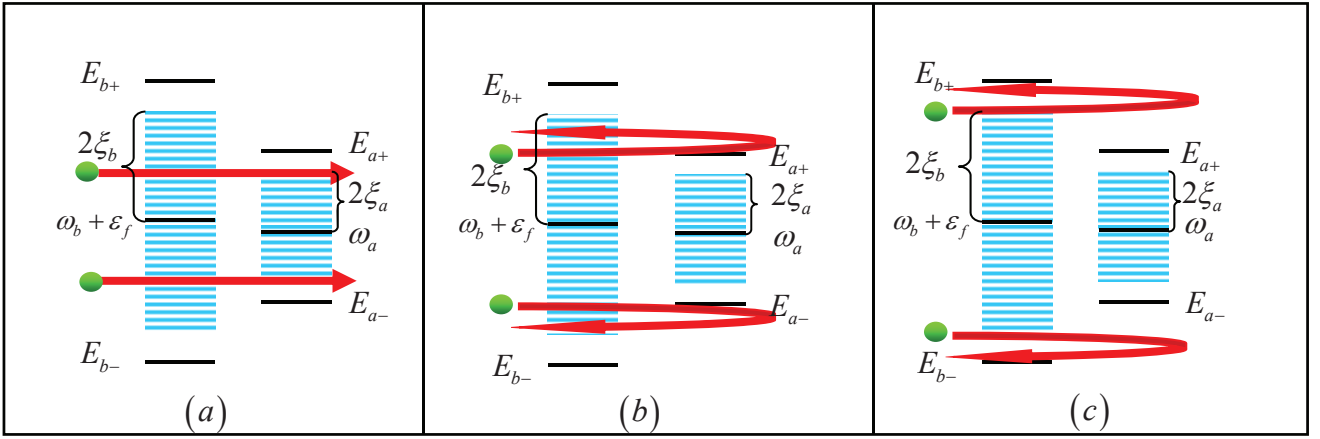} 
\end{center}
\caption{(Color online) Schematic illustration of single photon energy
spectrums in the chain \textit{A} and chain \textit{B}. Three kinds of
nontrivial cases: (a) The photon incident energy resonate with continuous
energy spectrums boundaries of the chain \textit{A}. (b) The photon incident
energy resonate with single photon bound state energy in Chain \textit{A}. (c)
The photon incident energy resonate with continuous energy spectrums
boundaries of the chain \textit{B}.}
\label{fig3}
\end{figure*}

In Fig.~\ref{fig2}, we plot the photon transmission rate $T=\left\vert
s\right\vert ^{2}$, reflection rate $R=\left\vert r\right\vert ^{2}$ and $%
\left\vert s\right\vert ^{2}+\left\vert r\right\vert ^{2}$ as a function of
photon energy $E$, from which we find the perfect transmission and
reflection occur at certain resonant points. Firstly, the transmission
generally vanishes at the energy band boundaries of chain \textit{B} with $%
k=0,\pi $. It follows from Eq.~(\ref{s}) that these energies are zeros of s.
Secondly, the photon is totally reflected when the incident energy equals to
the energy band boundaries of chain \textit{A} which correspond to $k^{\prime
}=0,\pi $. Thirdly, the energies of single photon bound states in chain
\textit{A} can be obtained by solving the transcendental equation

\begin{equation}
E=\varepsilon _{e}\pm J_{a}^{2}/\sqrt{\left( E-\omega _{a}\right) ^{2}-4\xi
_{a}^{2}}
\end{equation}%
which gives $E=0.68$ and $E=1.3$ by choosing parameters as $\varepsilon
_{e}=0.95$, $\omega _{a}=1$, $J_{a}=0.15$, and $\xi _{a}=0.15$. When the
photon incidents with the energy accidentally resonating with the bound
states energies, the perfect reflection takes place. This is just the
Fano-Feshbach resonate phenomenon. In addition, when energy of the photon is
at the range $\omega _{a}-2\xi _{a}<E<\omega _{a}+2\xi _{a}$, the summation
of the photon transmission rate and reflection rate is less than $1$ as
shown in Fig.~\ref{fig2} since incident photon in chain \textit{B} can be
scattered into chain \textit{A}. In Fig.~\ref{fig3}, we plot the energy
spectrum sketch map of our system and display three kinds of nontrivial
cases for $R=1$ and $T=1$.

Considering the improvement of experimental technical in solid state system, our model can be realized in several physical systems of superconducting qubits interacting with microwave stripline resonators\cite{jma}, quantum dot with photonic crystal defects\cite{khe}, as well as a nature atom with monolithic microresonator\cite{tao}. Especially it is reported that in a single quantum dot and semiconductor microcavity system, the coupling strength is about 70$\mu ev$\cite{ama} and the coupling strength between cavities can achieve 0.8THz\cite{jpr}. These parameters are suitable to our consideration in the above discussion.

\section{Dark state mechanism}

\label{sec5}

In the intersectional resonator, the $\Lambda $-type atom couples to two
field modes of the two CRAs, so that we can realize EIT effect by tuning one
of the field modes. In the present section we will explore the coherent
control the single photon transfer in our system.

The effective Hamiltonian in the interaction picture is obtained as

\begin{eqnarray}
H_{I}^{\prime } &=&e^{iH_{0}^{\prime }t}H_{I}e^{-iH_{0}^{\prime
}t}-H_{0}^{\prime }  \notag \\
&=&-\Delta _{1}\left\vert e\right\rangle \left\langle e\right\vert -\left(
\Delta _{1}-\Delta _{2}\right) \left\vert f\right\rangle \left\langle
f\right\vert   \notag \\
&&+\left[ \left( J_{a}\left\vert e\right\rangle \left\langle g\right\vert
a+J_{b}\left\vert e\right\rangle \left\langle f\right\vert b\right) +h.c.%
\right]
\end{eqnarray}

We omit the subscript $0$ of the photon annihilation and creation operators
in the intersectional resonator. Here we choose%
\begin{eqnarray}
H_{0}^{\prime } &=&\varepsilon _{f}\left\vert f\right\rangle \left\langle
f\right\vert +\varepsilon _{e}\left\vert e\right\rangle \left\langle
e\right\vert +\omega _{a}a^{\dag }a+\omega _{b}b^{\dag }b  \notag \\
&&+\Delta _{1}\left( \left\vert e\right\rangle \left\langle e\right\vert
+\left\vert f\right\rangle \left\langle f\right\vert \right) -\Delta
_{2}\left\vert f\right\rangle \left\langle f\right\vert
\end{eqnarray}%
with $\Delta _{1}=\varepsilon _{e}-\omega _{a}$ and $\Delta _{2}=\varepsilon
_{e}-\varepsilon _{f}-\omega _{b}$ are detunings.

Under the two-photon resonant condition $\Delta _{1}=\Delta _{2}=\Delta $,
the interaction Hamiltonian is

\begin{equation}
H_{I}^{\prime }=-\Delta \left\vert e\right\rangle \left\langle e\right\vert
+ \left[ \left( J_{a}\left\vert e\right\rangle \left\langle g\right\vert
a+J_{b}\left\vert e\right\rangle \left\langle f\right\vert b\right) +h.c.%
\right] .
\end{equation}

The eigenvalues of the above Hamiltonian are

\begin{eqnarray}
E_{\pm } &=&-\frac{\Delta \pm J^{\prime }}{2}, \\
E_{0} &=&0.
\end{eqnarray}
With the corresponding eigenstates:

\begin{eqnarray}
\left\vert B_{\pm }\right\rangle &=&\frac{1}{\chi }\left[ \left( J^{\prime
}\mp \Delta \right)\left\vert \phi,e\right\rangle \pm 2\left( J_{a}\left\vert
0,g\right\rangle +J_{b}\left\vert 0,f\right\rangle \right) \right]  \notag \\
\left\vert D\right\rangle &=&\frac{1}{J}\left( J_{a}\left\vert
0,f\right\rangle -J_{b}\left\vert 0,g\right\rangle \right) ,
\end{eqnarray}
where $J=\sqrt{J_{a}^{2}+J_{b}^{2}}$, $J^{\prime }=\sqrt{4\left(
J_{a}^{2}+J_{b}^{2}\right) +\Delta ^{2}}$, and $\chi =\sqrt{\left( J^{\prime
}-\Delta \right) ^{2}+4J^{2}}$.

The state $\left\vert D\right\rangle $ with vanishing eigen-energy is called
dark state because it does not evolve with time and does not transmit to
the excited state. The other two states $\left\vert B_{\pm }\right\rangle $
are called bright states.

We rewrite the total Hamiltonian $H$ in the basis of the state $\left\vert
B_{\pm}\right\rangle $, $\left\vert D\right\rangle $, $\left\vert
m,g\right\rangle $, $\left\vert n,f\right\rangle $ to include the free
Hamiltonian

\begin{eqnarray}
H_{free} &=&\left( \frac{4J^{2}-\Delta ^{2}}{2J^{\prime }}+\omega
_{a}+\omega _{e}\right) \left\vert B_{+}\right\rangle \left\langle
B_{+}\right\vert  \notag \\
&&+\left( \omega _{a}+\omega _{e}-\frac{1}{2}\right) \left\vert
B_{-}\right\rangle \left\langle B_{-}\right\vert +\omega _{a}\left\vert
D\right\rangle \left\langle D\right\vert  \notag \\
&&+\left( \omega _{b}+\epsilon _{f}\right) \left( \sum_{m\neq
0}a_{m}^{\dagger }a_{m}+\sum_{n\neq 0}b_{n}^{\dagger }b_{n}\right)
\label{free1}
\end{eqnarray}

and the coupling terms among these states

\begin{eqnarray}
H_{coup} &=&\frac{2\Delta J}{J^{\prime }}\left( \left\vert
B_{+}\right\rangle \left\langle B_{-}\right\vert +h.c.\right)  \notag \\
&&+\{[\left( \left\vert 1,f\right\rangle +\left\vert -1,f\right\rangle
-\left\vert 1,g\right\rangle -\left\vert -1,g\right\rangle \right)  \notag \\
&&\cdot \frac{\chi \left( J^{\prime }-\Delta \right) }{J^{\prime }}\left(
J_{a}\xi _{a}\left\langle B_{+}\right\vert +J_{b}\xi _{b}\left\langle
B_{-}\right\vert \right)  \notag \\
&&+\frac{1}{J}[J_{b}\xi _{a}\left( \left\vert 1,g\right\rangle +\left\vert
-1,g\right\rangle \right) \left\langle D\right\vert  \notag \\
&&-J_{a}\xi _{b}\left( \left\vert 1,f\right\rangle +\left\vert
-1,f\right\rangle \right) \left\langle D\right\vert ]+h.c.\}  \notag \\
&&-\xi _{a}\sum_{m}\left( a_{m}^{\dagger }a_{m+1}+h.c.\right)  \notag \\
&&-\xi _{b}\sum_{n}\left( b_{n}^{\dagger }b_{n+1}+h.c.\right) .
\label{transition}
\end{eqnarray}

\begin{figure}[tbp]
\begin{center}
\includegraphics[bb=130 280 445 733,width=8cm]{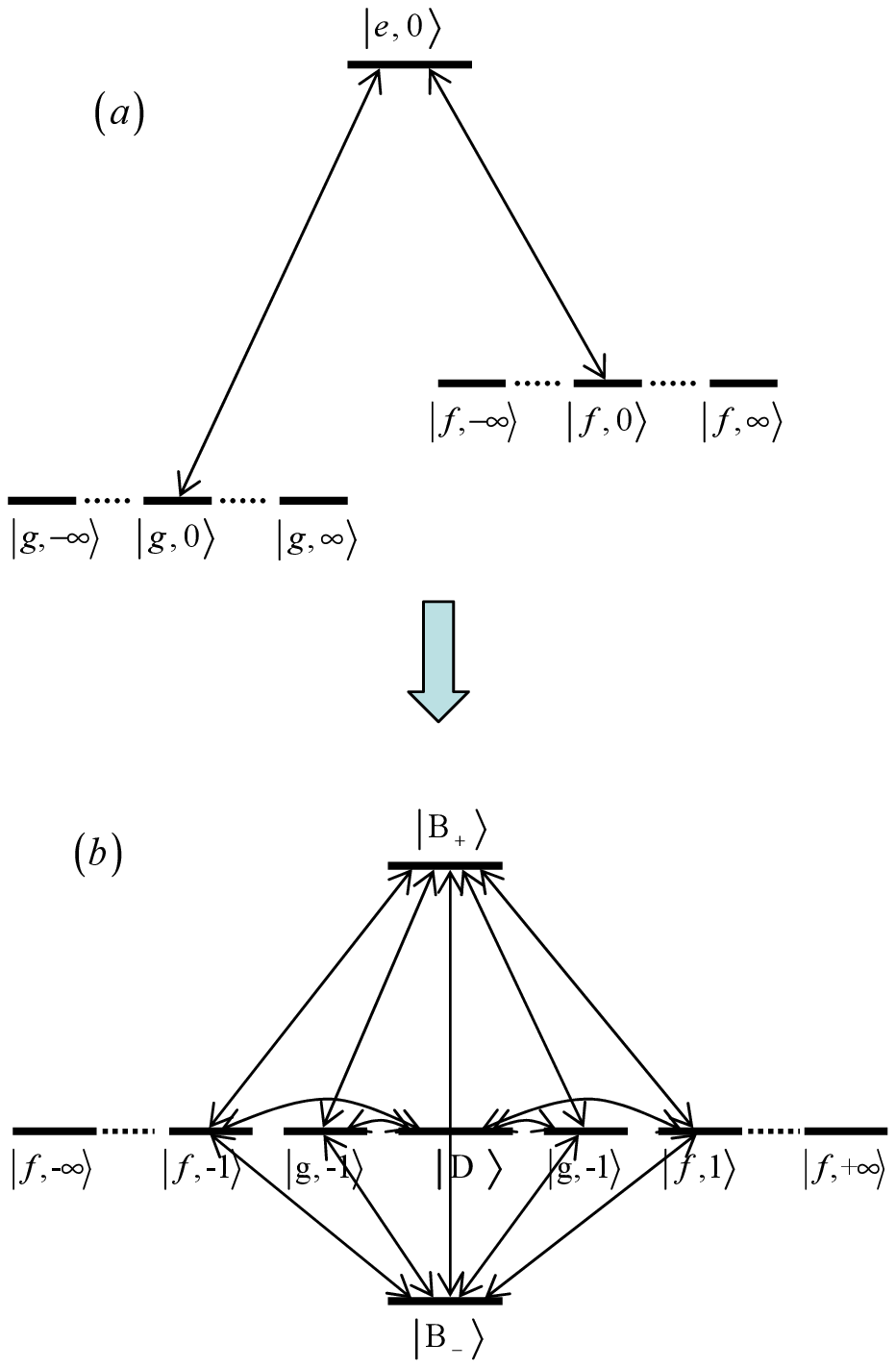}
\vspace{0.5cm}
\end{center}
\caption{(Color online) (a)Level coupling scheme under the original
presentation. (b)Level coupling scheme under the dark state presentation.}
\label{fig7}
\end{figure}

In the intersectional resonator, the couplings between the single excitation
states are displayed in Fig.~\ref{fig7}(a). Diagram illustrating of coherent
interactions mediated by the dark state and two bright states are also shown
in Fig.~\ref{fig7}(b).

In order to study the effect of the dark state and bright states channels in
single photons transmission, we adjust the system parameters to satisfy the
two photon resonate condition $\omega _{a}=\omega _{b}+\varepsilon _{f}$. It
can be seen from Eq.~(\ref{free1}) that the expectation value of the
Hamiltonian for dark state is $\omega _{a}$, which is equal to that of other
single excitation states outside the resonators. It means that photon
transmission through the dark state channel is easier than the other two
bright channels. To explore the different roles of dark state channel and
two bright states channels for the photon transmission, we calculate the
overlap between the single photon energy eigenstate and these three states,
which are expressed as

\begin{eqnarray}
\langle E|D\rangle &=&\frac{1}{J}\left[ J_{a}u_{f}^{\ast }\left( 0\right)
-J_{b}u_{g}^{\ast }\left( 0\right) \right] , \\
\langle E|B_{+}\rangle &=&\frac{1}{\chi }\left[ 2J_{a}u_{g}^{\ast }\left(
0\right) +2J_{b}u_{f}^{\ast }\left( 0\right) +\left( J^{\prime }-\Delta
\right) u_{e}^{\ast }\right] ,  \notag \\
&& \\
\langle E|B_{-}\rangle &=&-\frac{1}{\eta }\left[ 2J_{a}u_{g}^{\ast }\left(
0\right) +2J_{b}u_{f}^{\ast }\left( 0\right) -\left( \Delta +J^{\prime
}\right) u_{e}^{\ast }\right] .  \notag \\
&&
\end{eqnarray}%
with
\begin{equation}
\eta =\sqrt{\left( \Delta +J^{\prime }\right) ^{2}+4J^{2}}.
\end{equation}%
Here $u_{g}\left( 0\right) $ and $u_{f}\left( 0\right) $ are defined in Eq.$%
\left( \ref{jie2}\right) $ and Eq.$\left( \ref{jie1}\right) $.

\begin{figure}[tbp]
\resizebox{0.45\textwidth}{!}{
\includegraphics{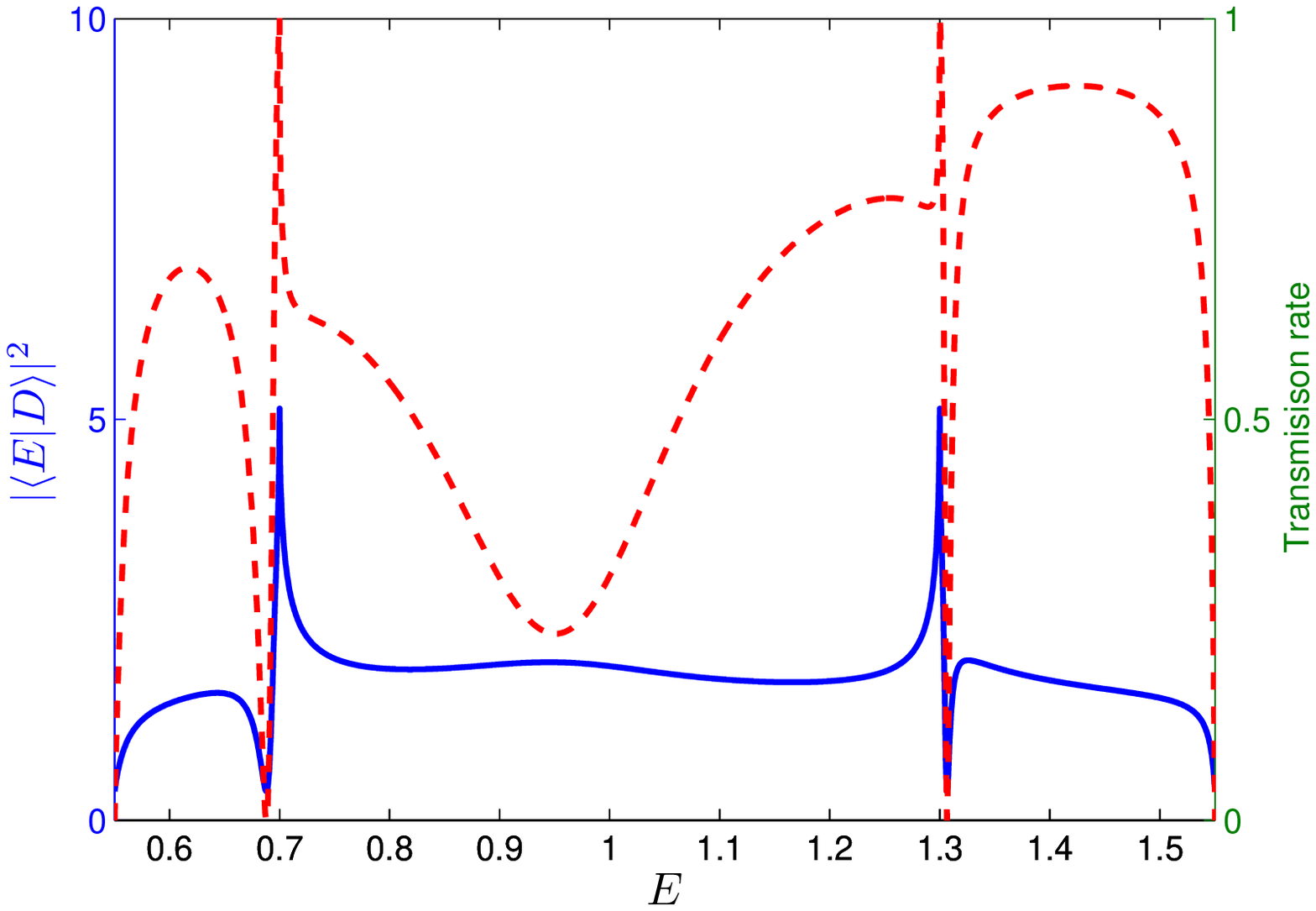}
}
\caption{(Color online) Norm of the overlap between the scatter state $%
|E\rangle$ and the dark state $|D\rangle$. The dashed(red) line represents
the transmission rate of the incident photon. }
\label{fig8}
\end{figure}

\begin{figure}[tbp]
\resizebox{0.45\textwidth}{!}{
\includegraphics{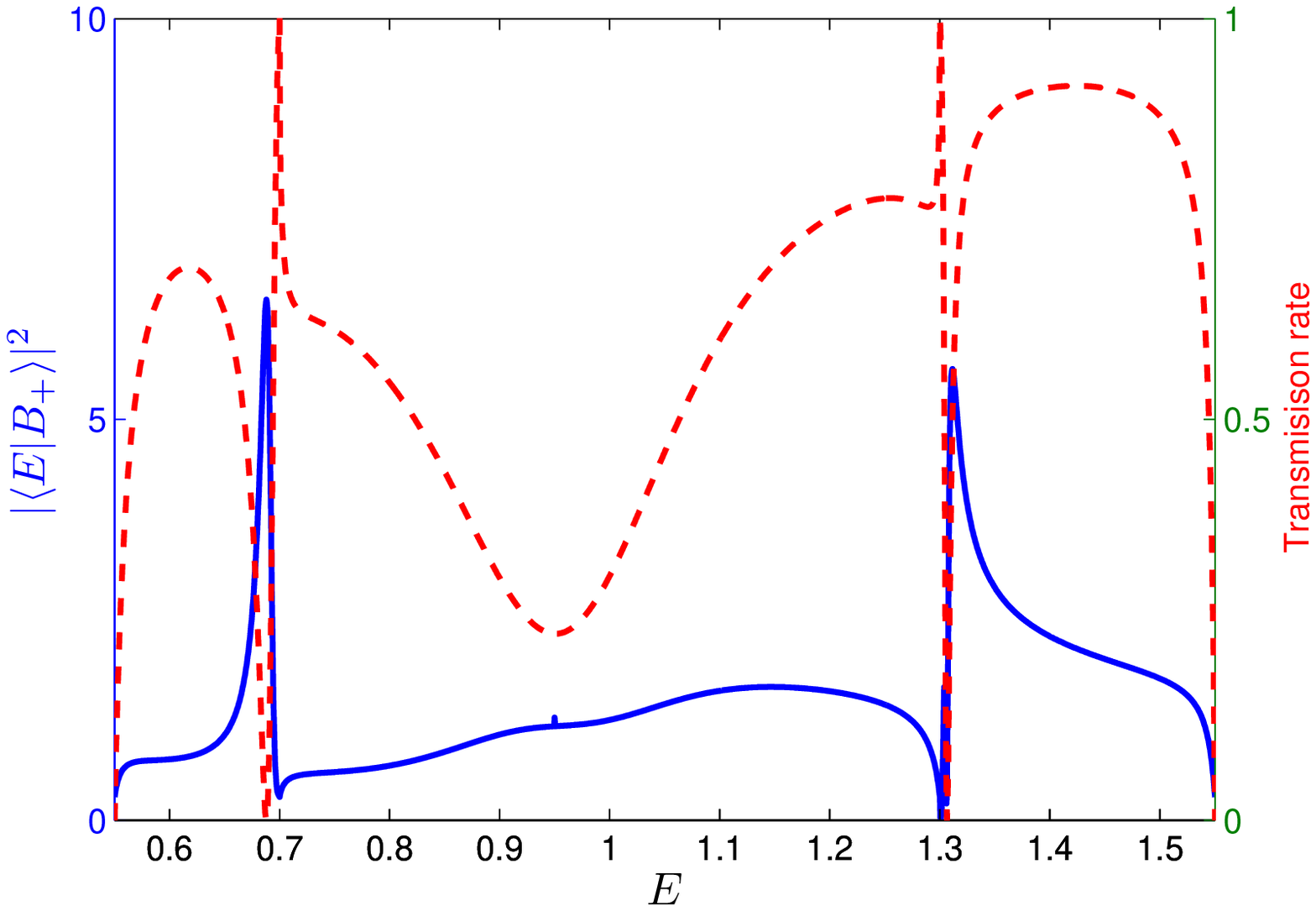}
}
\caption{(Color online) Norm of the overlap between the scatter state $%
|E\rangle$ and the bright state $|B_{+}\rangle$. The dashed(red) line
represents the transmission rate of the incident photon.}
\label{fig9}
\end{figure}

\begin{figure}[tbp]
\resizebox{0.45\textwidth}{!}{
\includegraphics{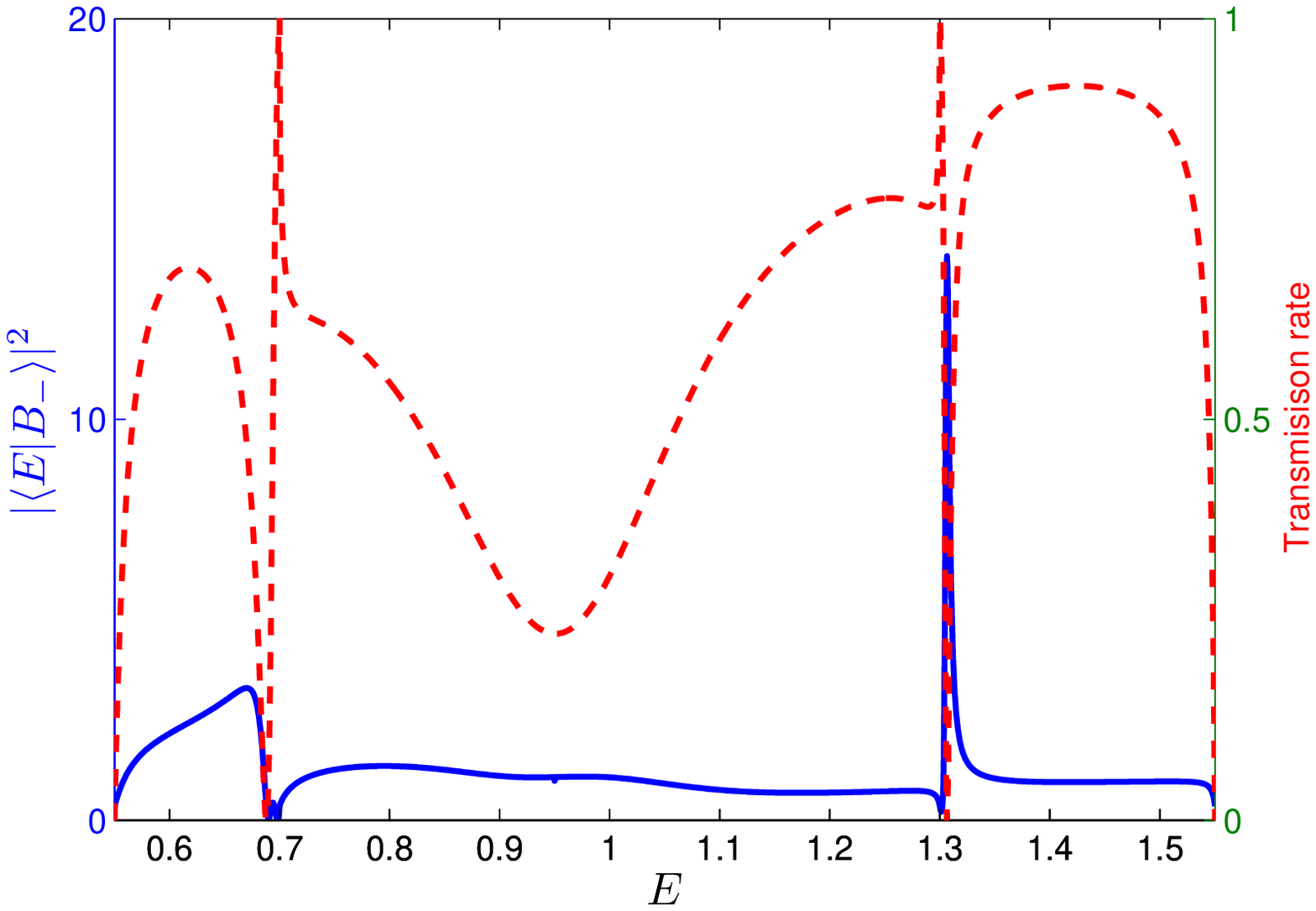}
}
\caption{(Color online) Norm of the overlap between the scatter state $%
|E\rangle$ and the bright state $|B_{-}\rangle$. The dashed(red) line
represents the transmission rate of the incident photon.}
\label{fig10}
\end{figure}

In Fig.~\ref{fig8}, we plot the norm of the overlap between the scattering state
and the dark state in the intersectional resonator. It is found that $%
|\langle E|D\rangle |^{2}$ has the similar shape as the transmission rate.
If the photon incident energy resonates with boundaries of the continuous
energy band of chain \textit{A}, the photon is totally transmitted. At these
two energies, $|\langle E|D\rangle |^{2}$ reaches its maximum value, namely
if the system is mainly populated in the dark state, the incident photon is
neither absorbed nor reflected by the atom in the intersectional resonator.
In this case we can reconstruct EIT effect in our system.

In Fig.~\ref{fig9} and Fig.~\ref{fig10}, we plot norm of the overlap between
the single excitation eigenstate and two bright states in the intersectional
resonator respectively. If the incident energy is resonate with single
photon bound state energies in chain \textit{A}, $|\langle E|B_{+}\rangle|^{2}$
and $|\langle E|B_{-}\rangle|^{2}$ reach their maximum values, where the incident
photon is totally reflected. Therefore, we conclude that the bright state
channels open when Fano-Feshbach resonate effect happens in our system.

\section{Conclusions and remarks}

\label{sec6}

In this paper we have studied the single photon coherent transfer in the
cross resonator arrays with a $\Lambda $-type atom which is localized in the
intersectional resonator. The coherent control of photon transfer can be
realized via the dark state mechanism with a fully quantum mechanism where no
classical field induce the EIT. It is shown that perfect reflection and
transmission can be realized when the photon incident energy resonate with
continuous energy spectrums boundaries in the two chains. There also exist
Fano-Feshbach resonance effect between the two resonator arrays. The
dark state mechanism in our system is also explored by considering the
condition to form EIT. In thus system we can coherently control single
photons transmission by using these properties. \newline

\noindent \textbf{Acknowledgments.}This work was supported by National
Natural Science Foundation of China under Grants No. 11121403, No. 10935010,
No. 11074261 and No. 11175044.

\section*{Appendix: Single photon bound state in the chain \textit{A}}

\label{Appendix} 

In this appendix we derive the single photons bound state energy in the
chain \textit{A}. Here we consider a two-level atom put into central resonator
of a coupled resonator array. The atom has ground state $|g\rangle$ and
excited state $|e\rangle$. We take the central resonator as the origin. The atom interact with the resonator field mode under the rotating wave approximation. The Hamiltonian of the system reads,

\begin{eqnarray}
H &=&H_{c}+H_{I}, \\
H_{c} &=&\omega _{a}\sum_{j}a_{j}^{\dagger }a_{j}-\xi _{a}\sum_{j}\left(
a_{j}^{\dagger }a_{j+1}+h.c.\right) , \\
H_{I} &=&\varepsilon \left\vert e\right\rangle \left\langle e\right\vert
+J_{a}\left( a_{0}\left\vert e\right\rangle \left\langle g\right\vert
+h.c.\right) .
\end{eqnarray}%
where $a_{j}^{\dagger }$ and $a_{j}$ are the creation and annihilation
operators of photon mode in the $j$th resonator with frequency $\omega _{a}$%
. $\xi _{a}$ is the hopping energy between nearest-neighbor resonators of
the field mode. We assume the energy of the atomic ground state is zero, $%
\varepsilon $ is the energy corresponding to excited state. $J_{a}$ is the
coupling strength between the 0th resonator field mode and the atom.

The stationary eigenstate of single excitation can be expressed as

\begin{equation}
\left\vert E\right\rangle =\sum_{j}u_{g}\left(j\right)\left\vert
1_{j},g\right\rangle +u_{e}\left\vert \phi,e\right\rangle.  \label{benzheng}
\end{equation}

Herein the state $\left\vert 1_{j},g\right\rangle $ corresponds to one
photon in the $jth$ resonator and the atom in its ground state, $\left\vert \phi,e\right\rangle$ corresponds to no photon in the resonator arrays and the
atom in its excited state.

The eigen equation $H\left\vert E\right\rangle =E\left\vert E\right\rangle $
results in the discrete stationary eigen equations

\begin{eqnarray}
\left(E-\omega_{a}\right)u_{g}\left(j\right)=&-&\xi_{a}\left[%
u_{g}\left(j+1\right)+u_{g}\left(j-1\right)\right]  \notag \\
&+&\frac{J_{a}^{2}u_{g}\left(0\right)}{E-\varepsilon}\delta_{j,0}.
\end{eqnarray}

The wave function of the bound state can be written as

\begin{equation}
u_{g}\left(j\right)=\left\{
\begin{array}{c}
Ae^{ikj},\text{ \ \ \ }j<0 \\
Ae^{-ikj}.\text{ \ \ }j>0%
\end{array}%
\right .
\end{equation}

Here $A$ is the normalized constant and $k$ is a complex number with negative
imaginary part.

At $j=0$, we can obtain the dispersion relation as

\begin{equation}
E=\omega_{a}-2\xi_{a}\cos k.
\end{equation}

At $j\neq0$,

\begin{equation}
E=\omega _{a}-2\xi _{a}e^{-ik}+\frac{J_{a}^{2}}{E-\varepsilon }.
\end{equation}

Then we can obtain the equation of bound state energy $E$ as

\begin{equation}
E=\varepsilon \pm \frac{J_{a}^{2}}{\sqrt{\left( E-\omega _{a}\right)
^{2}-4\xi _{a}^{2}}}.  \label{shufutainengl}
\end{equation}

\end{document}